\documentclass[12pt]{article}
\usepackage{cite,amssymb,graphicx,graphics}
\usepackage{amssymb}
\usepackage{amsmath}
\usepackage{color}

\begin{document}
\title{\bf Massless fermions in the standing wave braneworld}
\author{{\bf Merab Gogberashvili}\\
Andronikashvili Institute of Physics, \\
6 Tamarashvili St., Tbilisi 0177, Georgia \\
and \\
Javakhishvili State University, \\
3 Chavchavadze Ave., Tbilisi 0128, Georgia\\
{\sl E-mail: gogber@gmail.com} \\\\
{\bf Pavle Midodashvili}\\
Ilia State University, \\ 3/5 Kakutsa Cholokashvili Ave., Tbilisi 0162, Georgia\\
{\sl E-mail: pmidodashvili@yahoo.com} \\\\
{\bf Levan Midodashvili}\\
Gori University, \\ 53 Chavchavadze St., Gori 1400, Georgia \\
{\sl E-mail: levmid@hotmail.com}}
\maketitle
\begin{abstract}
We investigate the 5D massless fermionic fields within the standing wave braneworld model. We show that in the case of increasing warp factor there exist localized left spinor field zero modes on the brane, while right fermion wave functions are not normalizable.
\end{abstract}

\vskip 0.5cm

%%%%%%%%%%%%%%%%%%%%%%%%%%%%%%%%%%%%%%%%%%%%%%%%%%%%%%%%%%%%%%%%%%%

The brane models \cite{Hi,brane} has attracted a lot of interest recently with the aim of solving several open questions in modern physics. A key requirement for realizing the braneworld idea is that the various bulk fields be localized on the brane. For reasons of economy and avoidance of charge universality obstruction \cite{DuRuTi} one would like to have a universal gravitational trapping mechanism for all fields. However, there are difficulties to realize such mechanism with exponentially warped space-times. In the existing (1+4)-dimensional models spin $0$ and spin $2$ fields can be localized on the brane with the decreasing warp factor \cite{brane}, spin $1/2$ field can be localized with the increasing factor \cite{BaGa}, and spin $1$ fields are not localized at all \cite{Po}. For the case of (1+5)-dimensions it was found that spin $0$, spin $1$ and spin $2$ fields are localized on the brane with the decreasing warp factor and spin $1/2$ fields again are localized with the increasing factor \cite{Od}. There exist also 6D models with non-exponential warp factors that provide gravitational localization of all kind of bulk fields on the brane \cite{6D}, however, these models require introduction of unnatural sources.

To solve the localization problem recently we had proposed the standing wave braneworld model \cite{Wave}, which is generated by collective oscillations of gravitational and scalar phantom-like fields (similar to \cite{GMS}) in 5D bulk space-time. The metric of the model in the case of increasing warp factor has the form:
\begin{equation} \label{metric}
ds^2 = e^{2a|r|}\left( dt^2 - e^{u}dx^2 - e^{u}dy^2 - e^{-2u}dz^2 \right) - dr^2~.
\end{equation}
Here $a = \sqrt{\Lambda/6} > 0$, where $\Lambda$ is 5D cosmological constant, and
\begin{equation} \label{separation}
u(t,r) = B ~\sin (\omega t)~ \xi(r)~, ~~~~~ \xi(r) = e^{-2a|r|}Y_2\left( \frac{\omega}{a} e^{-a|r|} \right),
\end{equation}
where $B$ is a constant, $\omega$ denotes the oscillation frequency of the standing wave, and $Y_2$ is second order Bessel function of the second kind. The solution (\ref{metric}) describes the brane at $r = 0$, which undergoes anisotropic oscillations and sends waves into the 'sea' of phantom-like scalar field in the bulk.

The $r$-dependent factor $\xi(r)$ of the metric function (\ref{separation}) has finite number of zeros along the extra coordinate axis, which form the nodes of the standing wave. These nodes can be considered as 4D space-time 'islands', where the matter particles can be bound \cite{Wave}. One of nodes of the standing wave must be located at the position of the brane. This can be achieved by imposing the condition
\begin{equation}\label{MetricFunctionAtOrigin}
Y_2\left( \frac{\omega}{a} \right)=0
\end{equation}
fixing the value of the frequency of standing wave, $\omega$, in terms of the curvature scale $a$. In what follows we assume:
\begin{equation}\label{Omega-a}
\frac{\omega}{a}= Z_{1}\simeq 3.3842,
\end{equation}
where $Z_{n}$ denotes the $n$-th zero of the function $Y_{2}$.

The standing wave can provide localization of the matter particles with energies much smaller than $\omega$. One of the classical analogs of this mechanism is the so called optical lattice with standing electromagnetic waves \cite{Opt}. In the previous papers \cite{GMM} we had demonstrated the explicit localization of scalar, vector and tensor fields zero modes in standing wave braneworld with increasing warp factor. In this article we investigate the localization problem for massless fermions within this model.

%%%%%%%%%%%%%%%%%%%%%%%%%%%%%%%%%%%%%%%%%%%%%%%%%%%%%%%%%%%%%%%%%%%%

For Minkowskian $4\times 4$ gamma matrices ($\{ \gamma^\alpha, \gamma^\beta \} = 2\eta^{\alpha\beta}$, Greek indices, $\alpha, \beta, ...= t,x,y,z$, numerate 4D coordinates) we use the Weyl basis,
\begin{equation}\label{MinkowskianGammaMatrices}
\begin{array}{l}
\gamma ^t =~ \left( {\begin{array}{*{20}{c}}
0&I\\
I&0
\end{array}} \right),~~~
{\gamma ^i} = \left( {\begin{array}{*{20}{c}}
0&-\sigma^i\\
\sigma^i&0
\end{array}} \right),~~~
\gamma ^5 = i \gamma^t\gamma^x\gamma^y\gamma^z = \left( {\begin{array}{*{20}{c}}
I&0\\
0&-I
\end{array}} \right),
\end{array}
\end{equation}
where $I$ and $\sigma^i$ ($i = x,y,z$) denote the standard $2\times2$ unit and Pauli matrices respectively.

Let us recall that four-component columns represent fermions in 5D, and that 5D gamma matrices can be chosen as:
\begin{equation} \label{Gamma}
\Gamma^A = h_{\bar A}^A\Gamma^{\bar A}~,~~~~~~\Gamma^{\bar A} = \left(\gamma^t,\gamma^x,\gamma^y,\gamma^z,i\gamma^5 \right)~, ~~~~~
\left\{ \Gamma ^A,\Gamma ^B \right\} = 2g^{AB}~,
\end{equation}
where capital Latin indices, $A,B,...= t,x,y,z,r$, stand for 5D space-time coordinates and $\bar A,\bar B, ...$, refer to 5D local Lorentz (tangent) frame. So according to (\ref{Gamma}) the curved-space gamma matrices $\Gamma^A$ are related to Minkowskian ones (\ref{MinkowskianGammaMatrices}) as:
\begin{eqnarray}\label{GammaMatricesRelation}
\Gamma ^t &=& e^{ - a|r|}~\gamma ^t ~, \nonumber \\
\Gamma ^x &=& e^{ - a|r| - u/2}~ \gamma ^x~, \nonumber \\
\Gamma ^y &=& e^{ - a|r| - u/2}~\gamma ^y~, \\
\Gamma ^z &=&  e^{ - a|r| + u}~ \gamma ^z~, \nonumber \\
\Gamma ^r &=& i\gamma ^5~. \nonumber
\end{eqnarray}

The {\it f\"{u}nfbein} for our metric (\ref{metric}),
\begin{eqnarray}
h^{\bar A}_A &=& \left( e^{ a|r|},e^{ a|r|+ u/2},e^{ a|r| + u/2},e^{ a|r|-u},1 \right)~,\nonumber \\
h^{\bar AA} &=& g^{AB} h_B^{\bar A} = \left( e^{ - a|r|},-e^{ -a|r|- u/2},-e^{ -a|r| - u/2},-e^{ - a|r| + u},-1 \right)~, \\
h^A_{\bar A} &=& \eta_{\bar A \bar B} h^{\bar B A} = \left( e^{ - a|r|},e^{ -a|r|- u/2},e^{ -a|r| - u/2},e^{ - a|r| + u},1 \right)~,\nonumber \\
h_{\bar AA} &=& \eta_{\bar A \bar B} h^{\bar B}_A = \left( e^{ a|r|},-e^{ a|r|+ u/2},-e^{ a|r| + u/2},-e^{ a|r| - u},-1 \right)~, \nonumber
\end{eqnarray}
is introduced through the conventional definition:
\begin{equation}\label{VielbeinDefinition}
g_{AB}=\eta _{\bar A\bar B}h^{\bar A}_A h^{\bar B}_B~.
\end{equation}

The 5D Dirac action for free massless fermions can be written as:
\begin{equation}\label{SpinorAction}
S = \int d^5x \sqrt g ~i\overline \Psi \left(x^A\right) \Gamma ^MD_M \Psi \left(x^A\right)~,
\end{equation}
where the determinant $\sqrt g$ of the metric (\ref{metric}) is:
\begin{equation} \label{determinant}
\sqrt g = e^{4a|r|}~,
\end{equation}
and covariant derivatives are defined as follows:
\begin{equation}
D_A = \partial_A + \frac 14 \Omega_A^{\bar B \bar C} \Gamma_{\bar B} \Gamma_{\bar C}~.
\end{equation}
In the last expression $\Omega_M^{\bar M \bar N}$ denote the spin-connections:
\begin{eqnarray}\label{Spin-Connection}
\Omega_M^{\bar M \bar N} = - \Omega_M^{\bar N \bar M} = \frac12 \left[ h^{N\bar M}\left( \partial _M h_N^{\bar N} - \partial_N h_M^{\bar N} \right) - h^{N\bar N}\left( \partial_M h_N^{\bar M} - \partial _N h_M^{\bar M} \right) - \right. \nonumber \\
\left. - h_M^{\bar A} h^{P\bar M}h^{Q\bar N} \left( \partial_P h_{Q\bar A} - \partial_Q h_{P\bar A} \right) \right] ~.
\end{eqnarray}
The non-vanishing components of the spin-connection in the background (\ref{metric}) are:
\begin{eqnarray}\label{Spin-ConnectionComponents}
\Omega_t^{\bar t \bar r} &=& - \left( e^{a|r|} \right)'~, \nonumber \\
\Omega_x^{\bar x \bar r} &=& \Omega_y^{\bar y \bar r} = - \left( e^{a|r| + u/2 } \right)'~, \nonumber \\
\Omega_z^{\bar z \bar r} &=& - \left( e^{a|r| - u} \right)' ~,\\
\Omega_x^{\bar x \bar t} &=& \Omega_y^{\bar y \bar t} = \frac{\partial \left( e^{u/2}\right)}{\partial t}~, \nonumber \\
\Omega_z^{\bar z \bar t} &=& \frac{\partial \left(  e^{ - u}\right)}{\partial t}~, \nonumber
\end{eqnarray}
where primes denote derivatives with respect to the extra coordinate $r$.

The corresponding to (\ref{SpinorAction}) 5D Dirac equation reads:
\begin{equation}\label{SpinorEquation}
i\Gamma^AD_A\Psi = i\left( \Gamma ^\mu D_\mu + \Gamma ^rD_r \right)\Psi = 0~.
\end{equation}
For the bulk fermion field wave function we use the chiral decomposition:
\begin{equation}\label{Psi}
\Psi \left(x^\nu,r\right) = \psi_L \left(x^\nu\right) \lambda (r) + \psi_R \left(x^\nu\right) \rho (r)~,
\end{equation}
where $\lambda(r)$ and $\rho(r)$ are extra dimension factors of the left and right fermion wave functions respectively. We assume that 4D left and right Dirac spinors,
\begin{equation} \label{chiral}
\gamma^5 \psi_L = - \psi_L~, ~~~~~ \gamma^5 \psi_R = + \psi_R ~,
\end{equation}
correspond to zero mode wave functions, i.e. they satisfy free Dirac equations:
\begin{equation} \label{Dirac-free}
i\gamma^\mu \partial_\mu \psi_L = i\gamma^\mu \partial_\mu \psi_R = 0~.
\end{equation}

Apart from the massless states $\psi_L$ and $\psi_R$, the 5D Dirac equation also have solutions corresponding to massive fermions. In the single brane models the masses of the bounded massive states are typically of order of the energy scale $a$, characterizing the brane as a topological defect in higher-dimensional space-time. These states are very heavy and we do not consider them here.

The solutions of (\ref{Dirac-free}) in our representation (\ref{MinkowskianGammaMatrices}) can be written in the form:
\begin{eqnarray} \label{psi-free}
\psi_R \left(x^\nu\right) =
\left( \begin{array}{c}R\\0\end{array} \right)
e^{ - i( Et - p_xx - p_yy - p_zz )}~,\nonumber \\
\psi_L \left(x^\nu\right) =
\left( \begin{array}{c}0\\L\end{array} \right)
e^{ - i( Et - p_xx - p_yy - p_zz )}~,
\end{eqnarray}
where the constant 2-spinors $L$ and $R$ satisfy the relations:
\begin{equation} \label{relations}
\left(E + \sigma^ip_i\right)L = \left(E - \sigma^ip_i\right)R = 0~.
\end{equation}

When the frequency $\omega$ of standing waves in the background metric (\ref{metric}) is much larger than frequencies associated with the energies $E$ of the fermions on the brane,
\begin{equation}
\omega \gg E~,
\end{equation}
we can perform time averaging of the oscillatory functions in the Dirac equation (\ref{SpinorEquation}). Explicit expressions for time averages where found in \cite{GMM}:
\begin{equation}\label{CovariantDerivatives-and-TimeAverages}
\left\langle u \right\rangle  = \left\langle u' \right\rangle= \left\langle \frac{\partial u}{\partial t}\right\rangle  = \left\langle \frac{\partial \left(e^{ bu}\right)}{\partial t}\right\rangle   = 0~,~~~~~
\left\langle e^{bu} \right\rangle = I_0 \left( |bB|\xi(r)\right),
\end{equation}
where $b$ is a constant and $I_0$ is the modified Bessel function of zero order. Time averages of the Dirac operators $\left\langle i\Gamma^MD_M \right\rangle$ are:
\begin{eqnarray}\label{TimeAveragesOfDiracOperators}
\left\langle i\Gamma^tD_t \right\rangle &=& ie^{ -a|r|}~\gamma ^t\partial_t - \frac 12 a ~sgn(r)\gamma^5~, \nonumber \\
\left\langle i\Gamma^xD_x \right\rangle &=& ie^{ -a|r|}\left\langle e^{ -u/2} \right\rangle \gamma ^x\partial_x - \frac 12 a ~sgn(r)\gamma^5~, \nonumber \\
\left\langle i\Gamma^yD_y \right\rangle &=& ie^{ - a|r|}\left\langle e^{ -u/2} \right\rangle \gamma ^y\partial_y - \frac 12 a ~sgn(r)\gamma^5~, \\
\left\langle i\Gamma^zD_z \right\rangle &=& ie^{- a|r|}\left\langle e^{ u} \right\rangle \gamma ^z\partial_z - \frac 12 a ~sgn(r)\gamma^5~, \nonumber \\
\left\langle i\Gamma^rD_r \right\rangle &=& -\gamma ^5\partial_r~, \nonumber
\end{eqnarray}
where $sgn(r)$ is the sign function.

Now the equation (\ref{SpinorEquation}) can be written in the form:
\begin{equation}\label{SpinorEquation1}
i\left[{\gamma^t}{\partial _t} + \left\langle e^{ u/2} \right\rangle \left(\gamma ^x\partial _x + \gamma ^y\partial _y\right)  + \left\langle e^{-u} \right\rangle \gamma ^z\partial _z\right]\Psi = e^{a|r|}\gamma^5 \left[2a ~sgn(r) + \partial_r \right]\Psi~.
\end{equation}
Using the solutions of free equations (\ref{psi-free}) and the relations (\ref{relations}) it can be rewritten as the system:
\begin{equation}\label{L-R}
\left( \begin{array}{*{20}{c}}
-e^{a|r|}\left[2a ~sgn(r) + \partial_r \right]&\sigma^iP_i(r)\\
- \sigma^iP_i(r)&e^{a|r|}\left[2a ~sgn(r) + \partial_r \right]
\end{array} \right)\left( \begin{array}{*{10}{c}} \rho (r)R\\
\lambda (r)L\end{array} \right) = 0~,
\end{equation}
where we have introduced functions $P_i (r)$:
\begin{eqnarray} \label{P-i}
P_x (r) &=& \left(\left\langle e^{-u/2} \right\rangle -1\right) p_x = \left[I_0\left(|B|\xi(r)/2\right)-1\right] p_x~, \nonumber \\
P_y (r) &=& \left(\left\langle e^{-u/2} \right\rangle - 1 \right)p_y = \left[I_0\left(|B|\xi(r)/2\right)-1\right] p_y~, \nonumber \\
P_z (r) &=& \left(\left\langle e^{u} \right\rangle - 1 \right)p_z = \left[I_0\left(|B|\xi(r)\right)-1\right] p_z~, \\
P^2(r) &=& P_x^2 + P_y^2 + P_z^2~. \nonumber
\end{eqnarray}
These functions can be considered as the components of $P(r)$:
\begin{equation}\label{P}
P^2(r) = P_x^2 + P_y^2 + P_z^2~,
\end{equation}
which we call {\it '$r$-dependent momentum'}.

From the second equation of the system (\ref{L-R}) it is straightforward to find
\begin{equation} \label{rho=lambda}
\rho (r)R = e^{a|r|} \frac {\sigma^iP_i(r)}{P^2(r)}\left[2a ~sgn(r) + \partial_r \right]\lambda (r)L~.
\end{equation}
Inserting (\ref{rho=lambda}) into the first equation of (\ref{L-R}) and multiplying the result by $\sigma^iP_i$, we receive the second order differential equation for the function $\lambda (r)$:
\begin{equation}\label{L-Equation}
\lambda'' + \left[ 5a~ sgn( r ) - \frac{P'}{P}\right] \lambda' + \left[4a \delta (r) + 6a^2 - 2a~sgn(r)\frac{P'}{P} - P^2e^{ - 2a|r|} \right] \lambda = 0~.
\end{equation}

Now let us investigate this equation in the two limiting regions: far from and close to the brane.

Close to the brane, $r \to \pm 0 $, the {\it '$r$-dependent momentum'} (\ref{P-i}) behaves as:
\begin{equation}
\left. P (r)\right|_{r\to \pm 0} = A r^2 + O(r^3)~,
\end{equation}
where $A$ is constant, and the equation (\ref{L-Equation}) takes the following asymptotic form:
\begin{equation}
\lambda'' + \left[ 5a~ sgn( r ) - \frac{2}{r}\right] \lambda' + \left[4a \delta (r) + 6a^2 - \frac {4a}{r}~sgn(r) \right] \lambda = 0~.
\end{equation}
This equation has the unique nontrivial solution:
\begin{equation}\label{L-0}
\lambda (r)|_{r \to \pm 0} = C e^{ - 2a|r|}~,
\end{equation}
where $C$ is a constant.

As it follows from (\ref{rho=lambda}) and (\ref{L-0}), in our setup the right fermionic modes are absent on the brane:
\begin{equation}\label{rho-0}
\rho (r)|_{r \to \pm 0} = 0 ~.
\end{equation}
Such different behavior of the left and right massless fermions on the brane is not surprising, since in our model the effective mass term in (\ref{SpinorEquation1}) is of $\gamma^5$-type,
\begin{equation}
m(r) = 2a \gamma^5sgn(r) e^{a|r|}~,
\end{equation}
with the gap:
\begin{equation}
|m(r) - m(-r)| = 4a \gamma^5 e^{a|r|}~.
\end{equation}

In the second limiting region $r \to \pm \infty $ the function $P'/P$ vanishes and the equation (\ref{L-Equation}) obtains the asymptotic form:
\begin{equation}\label{L-infinity_0}
\lambda'' + 5a~ sgn( r ) \lambda' + 6a^2 \lambda = 0~,
\end{equation}
with the solution:
\begin{equation}\label{L-infinity}
\lambda (r)|_{r \to \pm \infty } \sim e^{ - 3a|r|}~.
\end{equation}
Using (\ref{L-infinity}) from the relation (\ref{rho=lambda}) we find the asymptotic behavior of the extra factor of the right fermion wave function:
\begin{equation}\label{R-infinity}
\rho (r)|_{r \to \pm \infty } \sim e^{ -2 a|r|}~.
\end{equation}

The condition for the localization of a field on the brane is that its wave function in the extra dimensions be normalizable, or that its action integral over $r$ be convergent. We had found that in our model the extra dimension part $\lambda (r)$ of the bulk left spinor wave function (\ref{Psi}) has the maximum at the origin,
\begin{equation}
\lambda (r)|_{r=0} = C~,
\end{equation}
falls off from the brane, and turns into the asymptotic form (\ref{L-infinity}) at the infinity. When $r\rightarrow \infty$ the determinant (\ref{determinant}) in the action integral for 5D spinor fields (\ref{SpinorAction}) increases as $e^{4a|r|}$. However, extra dimension factor of left fermions according to (\ref{L-infinity}) contribute $e^{-6a|r|}$ and overall $r$-depended part in (\ref{SpinorAction}) decreases as $e^{-2a|r|}$. So in the case of left fermions the integral over $r$ in (\ref{SpinorAction}) is convergent, i.e. zero modes of left fermions are localized on the brane.

At the same time according to (\ref{rho-0}) right spinor zero modes does not exist on the brane and due to (\ref{R-infinity}) at the infinity the extra dimension part of right fermions $\rho (r)$ decreases as $e^{ -2a|r|}$. It means that for right fermions integral over $r$ in (\ref{SpinorAction}) diverges and zero modes of right fermions are not normalizable on the brane.

%%%%%%%%%%%%%%%%%%%%%%%%%%%%%%%%%%%%%%%%%%%%%%%%%%%%%%%%%%%%%%%%%

To conclude in this latter we have studied localization problem of the massless fermions in the 5D standing wave braneworld model. We have found that in the case of increasing warp factor there exist left spinor field zero modes localized on the brane, while right fermionic wave functions are not normalizable.

\medskip

%%%%%%%%%%%%%%%%%%%%%%%%%%%%%%%%%%%%%%%%%%%%%%%%%%%%%%%%%%%%%%%%%

\noindent {\bf Acknowledgments:} The research was supported by the grant of Shota Rustaveli National Science Foundation $\#{\rm GNSF/ST}09\_798\_4-100$.

%%%%%%%%%%%%%%%%%%%%%%%%%%%%%%%%%%%%%%%%%%%%%%%%%%%%%%%%%%%%%%

\end{document}